\documentclass[twocolumn,showpacs,amsmath,amssymb]{revtex4}
\input{epsf}

\usepackage{graphicx}
\usepackage{array}
\usepackage{bigstrut}
\usepackage{longtable}
\usepackage{rotating,booktabs}
\usepackage{booktabs,threeparttable}
\usepackage{color}
\usepackage{bm}
\usepackage{footnote}

\begin{document}

\title{QED and relativistic nuclear recoil corrections to the 413 nm tune-out wavelength for the $2\,^3S_1$ state of helium}

\author{Yong-Hui Zhang,$^{1,\sharp}$ Fang-Fei Wu,$^{1,2,\sharp}$~\footnotetext{$\sharp$ Both authors contributed equally to this work} Pei-Pei Zhang,$^{1}$
Li-Yan \nolinebreak Tang,$^{1,*}$~\footnotetext{*Email Address: lytang@wipm.ac.cn} Jun-Yi Zhang,$^{1}$ K. G. H. Baldwin,$^3$ and Ting-Yun Shi$^{1,4,\dag}$~\footnotetext{$\dag$Email Address: tyshi@wipm.ac.cn}}

\affiliation {$^1$State Key Laboratory of Magnetic Resonance and
Atomic and Molecular Physics, Wuhan Institute of Physics and
Mathematics, Chinese Academy of Sciences, Wuhan 430071, People's Republic of China}

\affiliation {$^2$University of Chinese Academy of Sciences, Beijing 100049, People's Republic of China}

\affiliation {$^3$ Research School of Physics and Engineering, Australian National University, Canberra, Australian Capital Territory 0200, Australia}

\affiliation {$^4$ Center for Cold Atom Physics, Chinese Academy of Sciences, Wuhan 430071, People¡¯s Republic of China}

\date{\today}

\begin{abstract}
Comparison of high accuracy calculations with precision measurement of the 413 nm tune-out wavelength of the He($2\,^3S_1$) state provides a unique test of quantum electro-dynamics (QED).
We perform large-scale relativistic-configuration-interaction (RCI) calculations of the tune-out wavelength, that include the mass-shift operator, and fully account for leading relativistic nuclear recoil terms in the Dirac-Coulomb-Breit (DCB) Hamiltonian. We obtain the QED correction to the tune-out wavelength using perturbation theory, and the effect of finite nuclear size is also evaluated. The resulting tune-out wavelengths for the $2\,^3S_1(M_J=0)$ and $2\,^3S_1(M_J=\pm 1)$ states are 413.084 26(4) nm and 413.090 15(4) nm, respectively. Incorporating the retardation correction of 0.000 560 0236 nm obtained by Drake {\em et al.} to compare with the only current experimental value of 413.0938(9stat)(20syst) nm for the $2\,^3S_1(M_J=\pm 1)$ state, there is 1.4$\sigma$ discrepancy between between theory and experiment, which stimulates further theoretical and higher-precision experimental investigations on the 413 nm tune-out wavelength. In addition, we also determine the QED correction for the static dipole polarizability of the He($2\,^3S_1$) state to be 22.5 ppm, which may enable a new test of QED in the future.
\end{abstract}

 \maketitle

Bound-state quantum electrodynamics (QED) is one of the most successful theories in modern physics, having been tested through precision measurement over a diverse spectrum of experimental realisations. For example, measurement of the bound-state $g$ factor in the hydrogen-like $^{28}$Si$^{13+}$ and $^{12}$C$^{5+}$ at the sub-ppb level~\cite{sturm11a,sturm13,sturm14} has provided one of the strictest QED tests.

In order to test QED theory in many-electron systems, calculations and measurements for helium, the simplest multi-electron atom, are of great importance. Measurements of the fine-structure splitting in the $2\,^3P$ manifold have yielded a test of QED predictions with a precision at the sub-ppb (10$^{-9}$) level~\cite{smiciklas10, zheng17, drake02b, pachucki10}. The Lamb shift of the $2\,^1S_0$ and $2\,^3S_1$ states has been determined, respectively, using the $2\,^1S_0\rightarrow 3\,^1D_2$~\cite{huang18a} and $2\,^3S_1\rightarrow 2\,^3D_1$ two-photon transitions~\cite{dorrer97a}. However, 4 standard deviations in the discrepancy between measurements for the helium nuclear charge radius, which are determined by two different methods (the $2\,^3S\rightarrow2\,^3P$~\cite{shiner95, pastor12, zheng17a} and $2\,^3S\rightarrow2\,^1S$~\cite{rooij11a, wim18a} transition frequencies combined with calculations of the QED and recoil corrections~\cite{yerokhin10a, patkos16, patkos17}), pose significant challenges to QED theory.

QED tests that do not rely on energy level determinations can potentially provide important independent verification, such as the experimental and theoretical determination of transition rates, but these are both inherently difficult and of much lower precision~\cite{dall08, hodgman09, hodgman09b}. Therefore, further experiments probing other non-energy properties of helium are important to deliver an independent validation of QED, provided that the corresponding progress in theory can be achieved.

QED contributions play an important role in the atomic polarizability of helium. The most accurate theoretical calculation of the ground-state static dipole polarizability of helium has now reached an accuracy of 0.2 ppm~\cite{lach04a}, which provides a non-energy QED test when compared with high-precision experimental measurements~\cite{schmidt07a,gaiser18a}. It is difficult to further improve this experimental accuracy, since a measurement of polarizability depends on precisely measuring the electric field strength.

However, the same QED effects are also reflected in the dynamic polarizability~\cite{piszczatowski15, puchalski16a}. The 413 nm tune-out wavelength for the He($2\,^3S_1$) state, where the dynamic polarizability equals to zero, provides a further non-energy scheme to test QED~\cite{mitroy13b}. Since the position of the tune-out wavelength does not depend on the details of the laser power or beam profile, a measurement of the tune-out wavelength can potentially achieve higher sensitivity to test QED calculations than a measurement of the static dipole polarizability.

This application of the 413 nm tune-out wavelength of metastable helium to test QED theory has sparked great interest in high-precision measurement and high-accuracy calculations~\cite{henson15,zhang15,zhang16,manalo17,drake18,drake18a}. The first hybrid calculations were carried out by Mitroy and Tang~\cite{mitroy13b}. In 2015, Henson {\em et al.}~\cite{henson15} performed the first experimental measurement utilizing a novel, highly sensitive technique, and reported a value of 413.0938(9stat)(20syst) nm ($\sim$5 ppm accuracy) for the $2\,^3S_1(M_J=1)$ state of $^4$He, two orders of magnitude more precise than the value of 413.02(9) nm first predicted in Ref.~\cite{mitroy13b}. Recently, Zhang {\em et al.} performed an ab-initio calculation of the tune-out wavelength by extending non-relativistic and relativistic configuration interaction (NRCI and RCI) methods~\cite{zhang15,zhang16}. The RCI value of 413.085 9(4) nm, which includes the finite nuclear mass and relativistic corrections, reduced the discrepancy between the theoretical value and measurement result from 134 ppm to 19 ppm. The remaining 19 ppm discrepancy was mainly due to neglected QED corrections, which provides motivation for the more detailed QED and higher-order nuclear recoil investigations in the present work.

In this paper, we improve on previous B-spline RCI methods by self-consistently taking into account the nuclear recoil corrections in the Dirac-Coulomb-Breit (DCB) framework, and perform the QED correction with perturbation calculation. We obtain the individual contributions of the nuclear recoil effect, QED and finite nuclear size corrections to the 413 nm tune-out wavelength and static dipole polarizability of the $^4$He($2\,^3S_1$) state.
And for the first time, the uncertainty in the static dipole polarizability has achieved an accuracy of 0.1 ppm. The present values of the tune-out wavelength will set a benchmark for future measurements to seriously test QED calculations at a higher level of accuracy.

It is convenient to efficiently calculate dynamic polarizabilities at off-resonance frequencies using a power series expansion, such as employed in determining the ground-state polarizability at the He-Ne laser wavelength of helium~\cite{bhatia98a,piszczatowski15}. However since the 413 nm tune-out wavelength is located near the $2\,^3S_1\rightarrow 3\,^3P_J$ resonance line, the power series expansion cannot be used. In this work, we employ the sum-over-states method~\cite{mitroy13b,zhang16} to obtain dynamic dipole polarizablities, then extract the tune-out wavelength from making $\alpha_1(\omega)=0$. Under linear polarized light with laser frequency $\omega$, the dynamic dipole polarizability for a state with angular momentum $J$ and magnetic quantum number $M_J$ is
\begin{eqnarray}
\alpha_1(\omega)=\alpha_1^S(\omega)+\dfrac{3M_{J}^2-J(J+1)}{J(2J-1)}\alpha_1^T(\omega)
\,,\label{e1}
\end{eqnarray}
where $\alpha_1^S(\omega)$ and $\alpha_1^T(\omega)$ are, respectively, the scalar and tensor dipole polarizabilities~\cite{zhang16}.

In order to take account of the nuclear recoil corrections, the mass shift (MS) operator $H_{MS}$, which explicitly includes the non-relativistic and leading relativistic components, $H_{NRMS}$ and $H_{RMS}$~\cite{tupitsyn03} respectively, has been added directly into the DCB Hamiltonian,
\begin{eqnarray}
H = H_{DCB}+H_{MS}=H_{DCB}+H_{NRMS}+H_{RMS}
\end{eqnarray}
\begin{eqnarray}
H_{DCB}&=&\sum\limits_{i=1}^{2}\big[c\bm{\alpha}_i\cdot\bm{p}_i+\beta m_ec^2-\dfrac{Z}{r_i}\big]+\dfrac{1}{r_{12}}\nonumber \\
&-&\dfrac{1}{2r_{12}}\left[\bm{\alpha}_1\cdot\bm{\alpha}_2+\left(\bm{\alpha}_1\cdot\hat{\bm{r}}_{12}\right)
                                                             \left(\bm{\alpha}_2\cdot\hat{\bm{r}}_{12}\right)\right] \label{e3}\\
H_{NRMS}&=&\dfrac{1}{2m_0}\sum\limits_{i, j}^{2}\bm{p}_i\cdot\bm{p}_j \label{e4}\\
H_{RMS} &=&-\dfrac{1}{2m_0}\sum\limits_{i, j}^{2}\dfrac{\alpha Z}{r_i}\big[\bm{\alpha}_i+\dfrac{(\bm{\alpha}_i\cdot\bm{r}_i)\bm{r}_i}{r_i^2}\big]\cdot\bm{p}_j                                                            \label{e5}
\end{eqnarray}
where $c$ is the speed of light, $Z$ is the nuclear charge, $m_e$ is the mass of the electron, $\bm{\alpha}_i$ and $\beta$ are the $4\times4$ Dirac matrices, $\bm{p}_i$ is the momentum operator, $r_i$ represents the distance of the $i$-th electron from the nucleus, $\hat{\bm{r}}_{12}$ is the unit vector of the electron-electron distance $\bm{r}_{12}$, $\alpha$ is the fine structure constant, and $m_0=7294.2995361\,m_e$~\cite{mohr12} is the nuclear mass of $^4$He.

The wave function of helium for a state is expanded as a linear combination of the configuration-state wave functions. And the configuration-state wave functions $|\phi_{ij}(JM_J)\rangle$ are constructed by $a_{im_i}^+|0\rangle$ and $a_{jm_j}^+|0\rangle$ with the angular quantum numbers $\ell_i$ and $\ell_j$ less than the maximum number of partial wave $\ell_{max}$,
\begin{eqnarray}
|\phi_{ij}(JM_J)\rangle=\eta_{ij}\sum\limits_{m_im_j}
\langle j_im_i;j_jm_j|JM_J\rangle
a_{im_i}^+a_{jm_j}^+|0\rangle
\, , \label{e6}
\end{eqnarray}
where $\eta_{ij}$ is a normalization constant, $\langle j_im_i;j_jm_j|JM_J\rangle$ represents the Clebsch-Gordan coefficient of $jj$ coupling, $|0\rangle$ is the vacuum state, $a_{im_i}^+|0\rangle$ represents the $i$-th single-electron wavefunction, which can be obtained by solving the single-electron Dirac equation using the Notre Dame basis sets~\cite{johnson88a} of $N$ number of B-spline functions with order of $k=7$~\cite{bachau01a}.

QED corrections to polarizability and tune-out wavelength are obtained by the perturbation theory using accurate energies and wavefunctions of previous NRCI calculations~\cite{zhang15}. According to the calculation of QED correction to static polarizability~\cite{pachucki00a}, the followed expression of QED corrections to the dynamic dipole polarizability can be derived,
\begin{widetext}
\begin{eqnarray}
\delta \alpha_1^{QED}(\omega)&=&2\Bigg[
\sum\limits_n\dfrac{\langle g|D|n\rangle\langle n|D|g\rangle\langle g|\delta H_{QED}|g\rangle[(E_n-E_g)^2+\omega^2]}{[(E_n-E_g)^2-\omega^2]^2}-2\sum\limits_{nm}\dfrac{\langle g|D|n\rangle\langle n|D|m\rangle\langle m|\delta H_{QED}|g\rangle(E_n-E_g)}{[(E_n-E_g)^2-\omega^2](E_m-E_g)}  \nonumber \\
&-&\sum\limits_{nm}\dfrac{\langle g|D|n\rangle\langle n|\delta H_{QED}|m\rangle\langle m|D|g\rangle[(E_n-E_g)(E_m-E_g)+\omega^2]}{[(E_n-E_g)^2-\omega^2][(E_m-E_g)^2-\omega^2]} \Bigg]
\,\label{e7}
\end{eqnarray}
\end{widetext}
where $|g\rangle$ represents the nonrelativistic wavefunction of the initial state, $|n\rangle$ and $|m\rangle$ represent nonrelativistic wavefunctions of intermediate states, $D$ is the electric dipole transition operator. The QED operator, $\delta H_{QED}=H_{QED}^{(3)}+H_{QED}^{(4)}$, expanded to $\alpha^3$- and $\alpha^4$-order for the He($2\,^3S$) state are defined respectively as~\cite{yerokhin10a}
\begin{eqnarray}
H_{QED}^{(3)}&=&\frac{4Z\alpha^3}{3}\left\{\dfrac{19}{30}+\ln\big[(Z\alpha)^{-2}\big]-\ln\big(\frac{k_0}{Z^2}\big)\right\}\nonumber \\
             &\times & \big[\delta^3({\bold{r}}_1)+\delta^3({\bold{r}}_2)\big]
              -\dfrac{14\alpha^3}{3} \big(\dfrac{1}{4\pi r_{12}^3}\big) \,,\label{e8}
\end{eqnarray}
\begin{eqnarray}
H_{QED}^{(4)}&=&\alpha^4\left\{\Big[ -\dfrac{9\zeta(3)}{4\pi^2}-\dfrac{2179}{648\pi^2}+\dfrac{3\ln(2)}{2}-\dfrac{10}{27}\Big]\pi Z \right.\nonumber\\
             &+& \left.\Big[ \dfrac{427}{96}-2\ln(2)\Big]\pi Z^2
               \right\} \big[\delta^3({\bold{r}}_1)+\delta^3({\bold{r}}_2)\big] \,, \label{e9}
\end{eqnarray}
where $\ln k_0$ is the Bethe logarithm and $\zeta(x)$ is the Riemann zeta function.

When an atom in external electric field $\cal{E}$, the Bethe logarithm involves the electric-field dependence term $\partial_{\varepsilon}^2 \ln k_0$, which introduces about 0.6\% of the total QED corrections to the ground-state polarizability~\cite{lach04a}. In our calculation, we use the value of $\ln k_0=$ 4.364 036 82(1)~\cite{drake99b} for a free atom. The correction from the electric-field derivative of Bethe logarithm is evaluated by indicating 1\% of the $\alpha^3$-order QED correction to the dynamic dipole polarizability.
The $\alpha^4$-order QED includes the one-loop and two-loop radiative effects. The nonradiative component is neglected since the contribution to helium $2\,^3S_1$ ionization energy from the nonradiative component accounts for less than 5\% of total $\alpha^4$-order QED correction~\cite{pachucki00b}. The Araki-Sucher correction (last term in Eq.~(\ref{e8})) contributes $-5.6\times 10^{-9}$ a.u. to helium $2\,^3S_1$ energy~\cite{zhangp15}, which is four orders of magnitude smaller than $1.67\times 10^{-5}$ a.u. from the first term of Eq.~(\ref{e8}), and two orders of magnitude smaller than the $\alpha^4$-order QED contribution of $2.91\times 10^{-7}$ a.u. So we omit the Araki-Sucher correction in the determination of the 413 nm tune-out wavelength.
\begin{table}[!htbp]
\caption{\label{t1} Convergence of the energy (in a.u.) for the $^4$He($2\,^3S_1$) state.}
\begin{ruledtabular}
\begin{tabular}{clll}
 \multicolumn{1}{c}{($\ell_{max}, N$)}
&\multicolumn{1}{c}{DCB}   &\multicolumn{1}{c}{DCB+NRMS}   &\multicolumn{1}{c}{DCB+MS} \\
\hline
(7, 40)  &--2.175 344 5653  &--2.175 045 2572   &--2.175 045 3806 \\
(8, 40)  &--2.175 344 5952  &--2.175 045 2851   &--2.175 045 4098 \\
(9, 40)  &--2.175 344 6132  &--2.175 045 3011   &--2.175 045 4224 \\
(10, 40) &--2.175 344 6157  &--2.175 045 3020   &--2.175 045 4282 \\
(10, 50) &--2.175 344 6220  &--2.175 045 3083   &--2.175 045 4270 \\
Extrap.  &--2.175 344 64(2) &--2.175 045 31(1)  &--2.175 045 43(1)  \\
Ref.~\cite{zhang16} &&--2.175 045 3(2)& \\
Ref.~\cite{yerokhin10a} &&&--2.175 045 451   \\
\end{tabular}
\end{ruledtabular}
\end{table}
\begin{table}[!htbp]
\caption{\label{t2} Convergence of the static dipole polarizability $\alpha_1(0)$ (in a.u.) and the tune-out wavelength $\lambda_t$ (in nm) for the $2\,^3S_1(M_J=0,\pm1)$ states of $^4$He.}
\begin{ruledtabular}
\begin{tabular}{cll}
 \multicolumn{1}{c}{($\ell_{max}, N$)}
&\multicolumn{1}{c}{$\alpha_1(0)(M_J=0)$}
&\multicolumn{1}{c}{$\alpha_1(0)(M_J=\pm1)$}\\
\hline
(7, 40)  &315.715 818 07 &315.724 122 42  \\
(8, 40)  &315.715 993 59 &315.724 290 09  \\
(9, 40)  &315.716 037 70 &315.724 343 39  \\
(10, 40) &315.716 053 51 &315.724 366 77  \\
(10, 50) &315.716 050 67 &315.724 377 89  \\
Extrap.  &315.716 05(1)  &315.724 38(1)   \\
Ref.~\cite{zhang16} &315.716 5(4) &315.724 8(4)  \\
\hline
 \multicolumn{1}{c}{($\ell_{max}, N$)}
&\multicolumn{1}{c}{$\lambda_t(M_J=0)$}
&\multicolumn{1}{c}{$\lambda_t(M_J=\pm1)$}\\
\hline
(7, 40)   &413.079 716 23 &413.085 585 95 \\
(8, 40)   &413.079 899 85 &413.085 764 03 \\
(9, 40)   &413.079 963 29 &413.085 832 75 \\
(10, 40)  &413.079 994 33 &413.085 867 66 \\
(10, 50)  &413.080 000 16 &413.085 882 02 \\
Extrap.   &413.080 00(1)  &413.085 89(1) \\
Ref.~\cite{zhang16}  &413.080 1(4)  & 413.085 9(4)\\
\end{tabular}
\end{ruledtabular}
\end{table}

The calculations of the nuclear recoil corrections on the energies, polarizabilities, and tune-out wavelengths are performed using our improved RCI method. Table~\ref{t1} gives a convergence test of the energy for the $2\,^3S_1$ state of $^4$He. The extrapolation was done by assuming that the ratio between two successive differences in energies stays constant as the $\ell_{max}$ and $N$ become infinitely large. The DCB energies in the second column don't include the nuclear recoil correction. The DCB+MS and DCB+NRMS columns present energies with and without relativistic nuclear recoil effects, respectively. Comparing the extrapolated results between DCB+MS and DCB+NRMS columns, it's found that the relativistic nuclear recoil effect of $H_{RMS}$ reduces $1.2\times 10^{-7}$ a.u. to the energy of the $2\,^3S_1$ state. The present DCB+NRMS value is in reasonable agreement with the previous RCI energy~\cite{zhang16}, where the relativistic nuclear recoil correction is not taken into account. Compared with the perturbation calculation~\cite{yerokhin10a}, which includes the leading $\alpha^2$-order relativistic correction, our DCB+MS energy agrees well with the result of $-$2.175 045 451 a.u. of Ref.~\cite{yerokhin10a}. The same energy accuracy for other $n\,^3S_1$ and $n\,^{1,3}P_{J}$ states with $n$ up to 8 is maintained in our calculations.
\begin{table}[!htbp]
\caption{\label{t3} Convergence of QED correction to the static dipole polarizability $\alpha_1(0)$ (in a.u.) and the 413 nm tune-out wavelength $\lambda_t$ (in nm) for the $^4$He($2\,^3S_1$) state. The number of $B$-splines $N=40$ is fixed. $\delta \alpha_1^{QED}(0)(\alpha^3)$ and $\delta \alpha_1^{QED}(0)(\alpha^4)$ represent the $\alpha^3$- and $\alpha^4$-order QED corrections to $\alpha_1(0)$ respectively. $\delta \lambda_t^{QED}(\alpha^3)$=$\lambda_t$(NRCI+$\alpha^3$~QED)$-\lambda_t$(NRCI) and $\delta\lambda_t^{QED}(\alpha^4)$=$\lambda_t$(NRCI+$\alpha^4$~QED)$-\lambda_t$(NRCI) represent the $\alpha^3$- and $\alpha^4$-order QED corrections to $\lambda_t$.}
\begin{ruledtabular}
\begin{tabular}{cll}
 \multicolumn{1}{c}{$\ell_{max}$}
&\multicolumn{1}{c}{$\delta \alpha_1^{QED}(0)(\alpha^3)$}
&\multicolumn{1}{c}{$\delta \alpha_1^{QED}(0)(\alpha^4)$}
\\
\hline
7   &0.006 899 132 62   &0.000 119 945 10   \\
8   &0.006 899 146 48   &0.000 119 945 35    \\
9   &0.006 899 152 88   &0.000 119 945 46  \\
10  &0.006 899 156 22   &0.000 119 945 52    \\
Extrap.   &0.006 899 158(2)   &0.000 119 946(1)    \\
\hline
\multicolumn{1}{c}{$\ell_{max}$}
&\multicolumn{1}{c}{$\delta \lambda_t^{QED}(\alpha^3)$}
&\multicolumn{1}{c}{$\delta \lambda_t^{QED}(\alpha^4)$}\\
\hline
7   &0.004 147 699 87   &0.000 072 114 31 \\
8   &0.004 147 716 05   &0.000 072 114 59 \\
9   &0.004 147 723 72   &0.000 072 114 72 \\
10  &0.004 147 727 74   &0.000 072 114 79 \\
Extrap.   &0.004 147 729(2)   &0.000 072 115(1) \\
\end{tabular}
\end{ruledtabular}
\end{table}

Table~\ref{t2} gives a convergence test of the static dipole polarizability and the 413 nm tune-out wavelength for the $^4$He($2\,^3S_1$) state. For $\alpha_1(0)$, present RCI values have seven convergent digits, which improves on previous RCI values~\cite{zhang16} by one order of magnitude. For $\lambda_t$, the convergence is very smooth as $\ell_{max}$ and $N$ increased. The tune-out wavelengths for the $2\,^3S_1(M_J=0)$ and $2\,^3S_1(M_J=\pm1)$ states are 413.080 00(1) nm and 413.085 89(1) nm, respectively. The present value of 413.085 89(1) nm is more accurate than the previous RCI result of 413.085 9(4) nm~\cite{zhang16} by one order of magnitude. The relativistic nuclear recoil correction decreases the tune-out wavelength by 0.02 picometer (pm).

Recently, Drake and Manalo carry out an independent calculation of the tune-out wavelength by solving the Schr$\ddot{o}$dinger equation with Hylleraas basis sets, the relativistic effects of relative $O(Z\alpha^2)$ are obtained by perturbation theory. They obtain the tune-out wavelength of 413.079 958(2) nm and 413.085 828(2) nm for the magnetic sublevel of $M_J=0$ and $M_J=\pm1$~\cite{drake18,drake18a}, respectively, which are in good agreement with our RCI values. It's worthy to mention that the method of Hylleraas coordinates allows accurate calculation of electron correlation effects, while present RCI calculations automatically include higher-order one-electron relativistic corrections and electron-electron correlation of relative order $Z \alpha^2$.

As pointed out in our previous paper~\cite{zhang16}, the main discrepancy between the earlier theory~\cite{zhang15} and experiment~\cite{henson15} for the 413 nm tune-out wavelength comes from omission of QED contributions to the theoretical value. In Table~\ref{t3}, we present the convergence test for the $\alpha^3$- and $\alpha^4$-order QED corrections to the static dipole polarizability and the 413 nm tune-out wavelength of $2\,^3S_1$ state. The numerical results of $\delta \alpha_1^{QED}(0)(\alpha^3)$ and $\delta \alpha_1^{QED}(0)(\alpha^4)$ converge fairly smoothly and monotonically to an extrapolated values of 0.006 895 171(1) a.u. and 0.000 119 876(1) a.u., with at least five converged digits. The $\alpha^3$-order QED correction contributes 4.147 729(2) pm to the tune-out wavelength, which is two orders of magnitude greater than the $\alpha^4$-order QED correction. The $\alpha^4$-order QED correction has four significant digits, which is more than satisfactory for our purposes.

In addition, we also evaluate the finite nuclear size effect on the static dipole polarizability and the tune-out wavelength by adopting the operator of $\frac{4\pi}{3}r_{\scriptscriptstyle{^4\textsl{He}}}^2[\delta^3({\bold{r}}_1)+\delta^3({\bold{r}}_2)]$~\cite{puchalski16a}, where $r_{\scriptscriptstyle{^4\textsl{He}}}=1.6755$ fm is the nuclear charge radius of $^4$He~\cite{angeli13}. The corrections due to finite nuclear size on $\alpha_1(0)$ and $\lambda_t$ are respectively, 4.58$\times 10^{-6}$ a.u. and 2.75 fm, which are negligible in the present work. But in the future, if a measurement of the 413 nm tune-out wavelength can reach $10^{-9}$ level of accuracy, it would have potential for the determination of the nuclear charge radius of helium, which is comparable with most of the precision spectroscopy methods~\cite{patkos16,patkos17,wim18a,zheng17a}.

\begin{table}[!htbp]
\caption{\label{t4} Contributions to the static dipole polarizability (in a.u.) and the 413 nm tune-out wavelength (in nm) for the $2\,^3S_1(M_J=0,\pm1)$ states of $^4$He.}
\begin{ruledtabular}
\begin{tabular}{llll}
 \multicolumn{1}{l}{Contribution}&\multicolumn{1}{c}{$M_J$}
&\multicolumn{1}{l}{$\alpha_1(0)$(a.u.)}
&\multicolumn{1}{l}{$\lambda_t$(nm)}\\
\hline
RCI + nuclear recoil  &0              &315.716 05(1)        &413.080 00(1)\\
RCI + nuclear recoil  &$\pm1$         &315.724 38(1)        &413.085 89(1)\\
$\alpha^3$ QED without $\partial_{\varepsilon}^2 \ln k_0$ &         &0.006 899 158(2)   &0.004 147 729(2) \\
$\alpha^4$ QED                                            &  &0.000 119 946(1)        &0.000 072 115(1) \\
$\alpha^3$ QED from $\partial_{\varepsilon}^2 \ln k_0$    &  &0.000 07(1)             &0.000 04(1)\\
Finite nuclear size     &                                    &0.000 004 58            &0.000 002 75 \\
Total                            &0                                              &315.723 14(4)        &413.084 26(4)\\
Total                            &$\pm1$                                          &315.731 47(4)        &413.090 15(4) \\
\end{tabular}
\end{ruledtabular}
\end{table}
\begin{figure}
\includegraphics[width=0.35\textwidth]{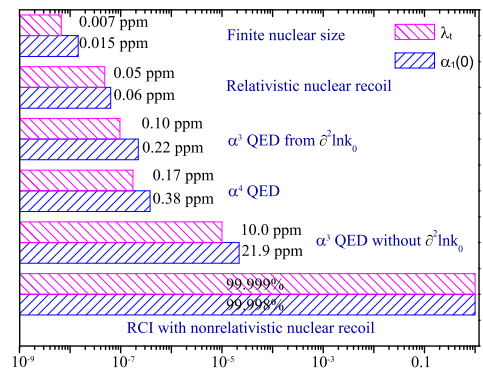}
\caption{\label{f1}(Color online) Relative contributions of various corrections to the static dipole polarizability $\alpha_1(0)$ and the tune-out wavelength $\lambda_t$ for the $2\,^3S_1(M_J=\pm 1)$ state of $^4$He.}
\end{figure}
\begin{figure}
\includegraphics[width=0.40\textwidth]{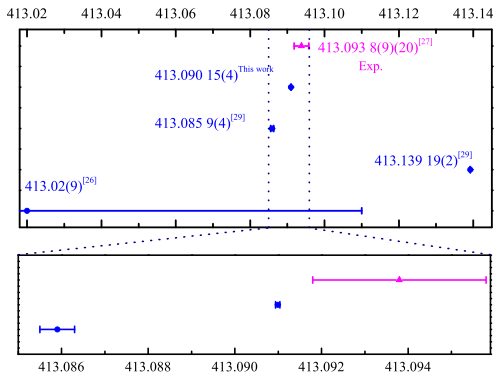}
\caption{\label{f3}(Color online) Comparisons of the tune-out wavelength $\lambda_t$ (in nm) for the $2\,^3S_1(M_J=\pm 1)$ state of $^4$He.}
\end{figure}

The individual and relative contributions from the QED, relativistic nuclear recoil, and finite nuclear size effects to static dipole polarizability and the 413 nm tune-out wavelength for the $2\,^3S_1(M_J=\pm1)$ state can be seen clearly from the Table~\ref{t4} and Fig.~\ref{f1}. The largest contribution to $\alpha_1(0)$ and $\lambda_t$ comes from the $\alpha^3$-order QED correction without $\partial_{\varepsilon}^2 \ln k_0$. The $\alpha^3$-QED correction from the electric-field dependence of the Bethe logarithm is hard to compute, but has been confirmed relatively small ($\sim0.6\%$) to the total QED correction in Ref.~\cite{puchalski16a,lach04a}. So in order to give a conservative estimation of this correction, we assume a 1$\%$ of the $\alpha^3$-order QED correction~\cite{puchalski16a} to reflect the contribution from $\partial_{\varepsilon}^2 \ln k_0$ term, which results in 0.000 07(1) a.u. correction to $\alpha_1(0)$. Combined with the $\alpha^3$- and $\alpha^4$-order QED corrections, the total QED contribution of 0.007 09(1) a.u. is added to the RCI values of 315.716 05(1) a.u. and 315.724 38(1) a.u., which gives 315.723 14(4) and 315.731 47(4) a.u. for the $2\,^3S_1(M_J=0)$ and $2\,^3S_1(M_J=\pm 1)$ states, respectively. The uncertainties, which are mainly from the $\partial_{\varepsilon}^2 \ln k_0$ term, have been doubled to be conservative. The total QED correction on the polarizability is 22.5 ppm. Like the ground-state polarizability, which has a similar QED contribution (22 ppm)~\cite{pachucki00a}, the contribution to the $2\,^3S_1$ state could also be measured as a test of QED.

For the 413 nm tune-out wavelength, seen from the Table~\ref{t4}, the $\alpha^3$-order QED correction without $\partial_{\varepsilon}^2 \ln k_0$ has about 10 ppm effect on $\lambda_t$. 1$\%$ of the $\alpha^3$-order QED correction is assumed to estimate the QED contribution from $\partial_{\varepsilon}^2 \ln k_0$ term. The total QED correction on the tune-out wavelength is then 0.004 26(1) nm. Adding this correction to our RCI values of 413.080 00(1) and 413.085 89(1) nm, we obtain the final tune-out wavelengths of 413.084 26(4) nm for $M_J=0$ and 413.090 15(4) nm for $M_J=\pm1$ magnetic sublevel of the $2\,^3S_1$ state, respectively. Comparison of calculations with measurement~\cite{henson15} is displayed in Fig.~\ref{f3}. The result of 413.085 9(4) nm~\cite{zhang16}, which does not includes the relativistic nuclear recoil and QED corrections, agrees with the measured value of 413.0938(9$_{stat}$)(20$_{syst}$) nm~\cite{henson15} at the level of 19 ppm. The present result of 413.090 15(4) nm for the $M_J=\pm1$ sublevel has included QED and relativistic nuclear recoil corrections. In order to make a meaningful comparison with the measurement~\cite{henson15} which probed the polarizability by using a traveling wave, the retardation correction to the tune-out wavelength needs to be taken into account. We incorporate Drake {\em et al.}'s retardation correction of 0.000 560 0236~\cite{drake19a} in our result of 413.090 15(4) nm to give 413.090 71(4) nm. Therefore a 1.4$\sigma$ discrepancy still exists in the tune-out wavelength between theory and experiment. So the present work provides considerable motivation for future experimental improvements to seriously test QED calculations at a higher level of accuracy.

In summary, we have calculated the dynamic dipole polarizability of the metastable helium under the DCB framework with the relativistic nuclear recoil effect included. The QED correction on the polarizability is taken into account using perturbation theory, and the finite nuclear size effect is also estimated. We precisely determine the tune-out wavelengths for the $^4$He($2\,^3S_1$) state for $M_J=0$ and $M_J=\pm1$ magnetic sublevels as 413.084 26(4) nm and 413.090 15(4) nm, respectively. We find that the relativistic nuclear recoil effect decreases the tune-out wavelength by $\sim$0.02 pm, and the QED corrections increase the tune-out wavelength by $\sim$4.26 pm. Our theoretical prediction for the 413 nm tune-out wavelength can be improved by introducing larger-scale configuration calculations with higher-order relativistic nuclear recoil effects included, and by calculating contributions from the field-dependent Bethe-logarithm in detail. We anticipate that this work will stimulate new high-precision measurements of the helium 413 nm tune-out wavelength to test QED calculations. In addition, we also obtained the static dipole polarizabilities for the $M_J=0$ and $M_J=\pm 1$ magnetic sublevels of the $^4$He($2\,^3S_1$) state as 315.723 14(4) a.u. and 315.731 47(4) a.u. respectively. We determined QED corrections for these polarizabilities of 22.5 ppm, which suggests that sensitive experimental measurements of static dipole polarizabilities of the $2\,^3S_1$ state might also be future test of QED.

We thank G. W. F. Drake and Z. C. Yan for useful discussion, and thank D. Cocks for helpful comments. This work was supported by the Strategic Priority Research Program of the Chinese Academy of Sciences, Grant Nos.XDB21010400 and XDB21030300, by the National Key Research and Development Program of China under Grant No.2017YFA0304402, and by the National Natural Science Foundation of China under Grants Nos.11474319, 11704398, 11774386, 11604369, and 91536102. L. Y. Tang and K. G. H. Baldwin acknowledge support from the Australian Research Council Discovery Project DP180101093.

%\bibliography{positron}

\end{document}